\documentclass{article}

    \PassOptionsToPackage{numbers, compress}{natbib}
    \usepackage[preprint]{neurips_2025}

\usepackage{microtype}
\usepackage{graphicx}
\usepackage{subfigure}
\usepackage{subcaption}
\usepackage{booktabs} % for professional tables

\usepackage{dsfont} % 引入dsfont包 
\newcommand{\indicator}[1]{\mathds{1}{#1}} % 定义指示函数的命令 

\usepackage{tabu}                     % 表格插入
\usepackage{multirow}                 % 一般用以设计表格，将所行合并
\usepackage{multicol}                 % 合并多列
\usepackage{multirow}                % 合并多行
\usepackage{float}                    % 图片浮动
\usepackage{makecell}                 % 三线表-竖线
\usepackage{booktabs}                 % 三线表-短细横线

\usepackage{xcolor}  % 导入颜色宏包

\usepackage{amsmath}
\usepackage{amssymb}
\usepackage{mathtools}
\usepackage{amsthm}
\usepackage{amsmath, amssymb}

\usepackage{algorithm}
\usepackage{algpseudocode}

\usepackage[utf8]{inputenc} % allow utf-8 input
\usepackage[T1]{fontenc}    % use 8-bit T1 fonts
\usepackage{hyperref}       % hyperlinks
\usepackage{url}            % simple URL typesetting
\usepackage{booktabs}       % professional-quality tables
\usepackage{amsfonts}       % blackboard math symbols
\usepackage{nicefrac}       % compact symbols for 1/2, etc.
\usepackage{microtype}      % microtypography
\usepackage{xcolor}         % colors

\usepackage[capitalize,noabbrev]{cleveref}

\title{Privacy Leaks by Adversaries: Adversarial Iterations for Membership Inference
Attack}

\author{%
  \AND
  Jing Xue\thanks{\texttt{xuejing0203@stu.xjtu.edu.cn}} \\
  \AND
  Zhishen Sun \\
  \And
  Haishan Ye \\
  \And
  Luo Luo \\
  \And
  Xiangyu Chang \\
  \And
  Ivor Tsang \\
  \And
  Guang Dai \\
}

\begin{document}

\maketitle

\begin{abstract}
Membership inference attack (MIA) has become one of the most widely used and effective methods for evaluating the privacy risks of machine learning models. These attacks aim to determine whether a specific sample is part of the model's training set by analyzing the model's output. While traditional membership inference attacks focus on leveraging the model’s posterior output, such as confidence on the target sample, we propose \texttt{IMIA}, a novel attack strategy that utilizes the process of generating adversarial samples to infer membership. We propose to infer the member properties of the target sample using the number of iterations required to generate its adversarial sample. We conduct experiments across multiple models and datasets, and our results demonstrate that the number of iterations for generating an adversarial sample is a reliable feature for membership inference, achieving strong performance both in black-box and white-box attack scenarios. This work provides a new perspective for evaluating model privacy and highlights the potential of adversarial example-based features for privacy leakage assessment. 
% Code is available at \href{https://anonymous.4open.science/r/Imia2-5477}{https://anonymous.4open.science/r/Imia2-5477}.
\end{abstract}

\section{Introduction}

Machine learning has widespread applications in many fields, such as autonomous driving \citep{aoki2023superdriverai, yu2024machine}, medical \citep{dixit2021risk} and financial systems \citep{chatzis2018forecasting,samitas2020machine}. Training a model requires collecting a large amount of data and aims to help the model learn knowledge that generalizes well from the training data through multiple epochs. For example, a hospital may train a diagnostic model using thousands of patients' CT scans and treatment outcomes. While, this model is used to assist in diagnosis and treatment, several studies \citep{carlini2022quantifying,salem2018ml,shokri2017membership} have shown that neural network models tend to remember their training data and an adversary can exploit this weakness to launch membership inference attack (MIA) \citep{carlini2022membership,choquette2021label,song2021systematic}. 
In such case, an adversary could infer whether a particular patient's record was used during the training process - potentially disclosing sensitive information like diagnosis results.

As a fundamental method to evaluate the privacy risk of machine learning models, membership inference attack (MIA) has received a lot of attention in recent years \citep{bertran2024scalable,debenedetti2024privacy,shokri2017membership,watson2021importance}. Specifically, given a target model, an adversary aims to know if a target sample was part of the model's training set (being a member) or not (being a non-member). Successful membership inference attack can reveal individual's health, consumption habits, or even home location \citep{carlini2021extracting}. Therefore, studying attack methods such as MIA is important for understanding and evaluating the privacy risk of machine learning models.

Membership inference attacks (MIA) typically fall into two categories. Distribution-based MIA methods rely on differences between the training and test data distributions but require large datasets and shadow models, making them resource-intensive \citep{carlini2022membership,chaudhari2023chameleon,tramer2022truth}. In contrast, metric-based MIA methods infer membership from the model’s output, such as confidence scores, without access to the training data or shadow models \citep{chen2022amplifying,song2019privacy,yeom2018privacy}. However, these methods are limited to scenarios where the model exposes soft outputs (e.g., probabilities). For instance, the Softmax Response attack is effective only when the target model outputs confidence values and fails when only hard labels are available.

These limitations prompt us to ask: whether there exists a universal method that can solve these limitations, and remain effective in black-box as well as white-box scenarios without requiring extensive data or computation.

In this paper, we affirmatively answer this question by proposing a novel membership inference attack method, Iterations for Membership Inference Attack (\texttt{IMIA}), from the lens of adversarial samples' generation. Our key observation is that member samples, being closer to the decision boundary, generally require more iterations to generate adversarial examples than non-member samples. \texttt{IMIA} leverages this iteration gap across different settings—including white-box and black-box—by employing suitable adversarial attack strategies such as HopSkipJumpAttack, SimBA, and PGD \citep{chen2020hopskipjumpattack,guo2019simple,madry2017towards}. Unlike prior work that relies on shadow models or access to similar data distributions \citep{carlini2022membership,tramer2022truth}, IMIA operates without requiring the training set, making it lightweight and broadly applicable.

In general, our contributions are summarized as follows:
\begin{enumerate}
    \item In this study, we propose a novel member inference attack method \texttt{IMIA} to infer whether the data belongs to the training set by analyzing the number of iterations required to generate adversarial samples. Different from traditional attack methods based on the posterior output of the target model, \texttt{IMIA} focuses on the generation process of adversarial samples, providing a tool to evaluate privacy leaks from the perspective of the internal operation of the model.
    \item \texttt{IMIA} does \emph{not} require any training data and training shadow models. 
    The target sample is sufficient for \texttt{IMIA} to execute the attack. 
    This strategy leverages the number of iterations required to generate adversarial samples from the target sample for MIA instead of posterior outputs of the target model.
    \item \texttt{IMIA} is highly adaptable and universal. Our proposed method IMIA can be exploited in all settings compared with previous methods that can only be used in one specific situation.
    We have conducted experiments on multiple models and datasets, covering different network architectures and data distributions. Though our proposed method is simple, experimental results show that our method can effectively evaluate the privacy leakage risk of the model under both black-box and white-box settings.
\end{enumerate}

\section{Background and Related Work}
In this section, we review research on membership inference attack, adversarial samples, and existing methods that we use as baselines.  

\subsection{Membership Inference Attack}
Membership inference attack has achieved great attention because it revealed that machine learning models have serious risks of privacy leakage and remember its training data \citep{bertran2024scalable,carlini2019secret,nasr2023scalable,prashanth2024recite,tobaben2024understanding}. In membership inference attack, given the target sample $x$, the adversary aims to infer whether this sample is in the training set $\mathrm {D_{tr}}$ of the target model $f_{\theta}$. As a result, membership inference attack can be seen as a binary classification privacy game. The participants in this game are the challenger $\mathcal{C}$ and the adversary $\mathcal{A}$. The game process can be broken down into several steps:
\begin{enumerate}
    \item The challenger samples a training set $\mathrm{D_{tr}\sim\mathbb{D}}$, and trains the target model $f_{\theta}$.
    \item The challenger randomly chooses a bit $b\in\{0,1\}$. If $b=0$, he will choose the target sample $(x,y)\in \mathbb{D}$ where $y$ is the ground-truth label of the target sample $x$, but $(x,y)$ is not in $\mathrm{D_{tr}}$. If $b=1$, he will choose the target sample in $\mathrm{D_{tr}}$ directly.
    \item The challenger sends the target sample to the adversary and allows the adversary to query the target model.
    \item The adversary gets the target sample and returns a bit $\hat{b}$ by querying the target model. If $\hat{b} = b$, the adversary will win this game.
\end{enumerate}
Based on the adversary's ability to access the target model, MIA can be divided into two categories:

\textbf{Black-box Membership Inference.} In the black-box setting, the adversary can only access the posterior output of the model (like confidence or hard labels) \citep{nasr2019comprehensive,wuyou}. This is a true scenario in the real world. In the black-box case, the attack methods can also be divided into two categories: the first is a distribution-based MIA \citep{carlini2022membership,chaudhari2023chameleon,debenedetti2024privacy,tramer2022truth}, in which the adversary needs to train a lot of additional shadow models as a proxy to mimic the target model. These shadow models require a large amount of data in the same distribution as the target sample and use the same model framework. The adversary trains shadow models with and without the target sample to learn the distribution difference between members and non-member samples, and then judges the member attributes of the target sample.

The second is the metric-based MIA, in which the adversary does not need to train additional shadow models. In Step~4, the adversary only uses the posterior output of the target model and designs a metric $\mathrm{Me}(\cdot)$, such as Softmax Response \citep{nasr2019comprehensive,song2019privacy}, Prediction Entropy \citep{salem2018ml,yeom2018privacy} and Modified Entropy \citep{song2021systematic}. Specifically, Softmax Response \citep{song2019privacy} computes the output probabilities of the target model $f_{\theta}(x)$ for the input sample $x$, obtaining the predicted probability $f_{\theta}(x)_i$ for each class $i$, and compares it with a preset threshold $\tau$. If the maximum predicted probability exceeds the threshold, the adversary infers that the sample belongs to the training set. Formally,
\begin{align*}
    I_{soft}(f_{\theta},(x,y))=\indicator\{\max f_{\theta}(x)_{i}\geq{\tau}\}
\end{align*}

\citet{salem2018ml} proposed to use the Prediction Entropy to conduct MIA. Prediction Entropy measures the uncertainty of the model's prediction and they thought that the prediction entropy of the member data would be close to 0 and a larger entropy value to the non-member data on the target model. This is formally expressed as:
\begin{align*}
    I_{ent}(f_{\theta},(x,y))=\indicator\{- \sum_i f_{\theta}(x)_i \log(f_{\theta}(x)_i)\leq{\tau}\}
\end{align*}
Besides, \citet{song2021systematic} proposed Modified Entropy that considered the ground-truth label of the target sample by decreasing the uncertainty on the true label of the target sample and increasing the uncertainty on the wrong label. It can be computed through:
\begin{align*}
Mentr(f_{\theta}(x),y)
=&
-(1-f_{\theta}(x)_y) \log(f_{\theta}(x)_y)\\
 -&\sum_{i\neq y} f_{\theta}(x)_i \log(1-f_{\theta}(x)_i)
\end{align*}
They inferred a target sample in $\mathrm{D_{tr}}$ according to:
\begin{align*}
    I_{Mentr}(f_{\theta},(x,y))=\indicator\{Mentr f_{\theta}(x,y) \leq{\tau}\}
\end{align*}
That is if the modified entropy value of the target sample is lower than the preset threshold, it will be recognized as a member.

\textbf{White-box Membership Inference.} In the white-box setting, the adversary can not only get the posterior output of the target model but also the internal parameters of the target model, like the loss and gradient information during the progress of the target model training \citep{carlini2017towards,goodfellow2014explaining}. \citet{yeom2018privacy} used the internal weights of the model and the loss of the target sample on the model to conduct MIA. If an adversary can get access to the logits output of the target model, he can conduct MIA through:

\begin{align*}
    I_{loss}(f_{\theta},(x,y))=\indicator\{- l (f_{\theta}(x), y)\leq{\tau}\}
\end{align*}

where the loss function will be cross-entropy loss. The sample $x$ which has lower loss value in the training set $\mathrm{D_{tr}}$. This capability allows an adversary to obtain more sensitive information of the victim model, but this capability is rarely available in the real world.

\subsection{Adversarial sample}
In a deep neural network, adversarial samples are inputs intentionally perturbed with small but deliberate perturbations that can cause deep neural networks to make incorrect predictions. 
\citet{goodfellow2014explaining} first revealed the vulnerability of neural networks to such inputs. 
Then, numerous attack methods have been proposed, including white-box approaches like PGD \citep{madry2017towards} and CW \citep{carlini2017towards}, as well as black-box approaches like SimBA \citep{guo2019simple}, which provide effective strategies for generating adversarial examples.

In the background of MIA, the characteristics of adversarial samples are used to infer the privacy of the model's training data. \citet{song2019privacy} used the confidence output on adversarial samples to judge the member attributes of the target samples. They believed that due to the robustness of the model, the adversarial samples generated from the member data show relatively stable prediction results on the model. Afterwards, \citet{choquette2021label} found that the distance from the adversarial sample to the model's decision boundary can be directly used to carry out MIA and had a great performance under the black-box setting. They judged a sample as a member if the distance from the adversarial sample to its decision boundary is larger than a preset threshold:
\begin{align*}
    I_{dis}(f_{\theta},(x,y))=\indicator\{d(x,\hat x) \geq{\tau}\}
\end{align*}
\citet{del2022leveraging} also analyzed MIA from this perspective.

In our paper, we carry out the metric-based black-box membership inference from the lens of the generation of adversarial samples. Different from prior works\citep{choquette2021label}, we do not measure the boundary distance between the adversarial sample and its decision boundary but record the number of iterations during the process of generating adversarial samples. Softmax Response, Prediction Entropy, and Modified Entropy as typical representatives in the score-based metric attacks, we choose these as our baselines. In the case where the target model only outputs hard labels, we choose the boundary distance attack as our baseline because boundary distance has a great performance in the metric-based condition.

\begin{figure*}[ht]
\vskip 0.2in
    \begin{minipage}[b]{0.48\linewidth}
        \centering
        \subfigure[a]{
    	    \includegraphics[width=\linewidth]{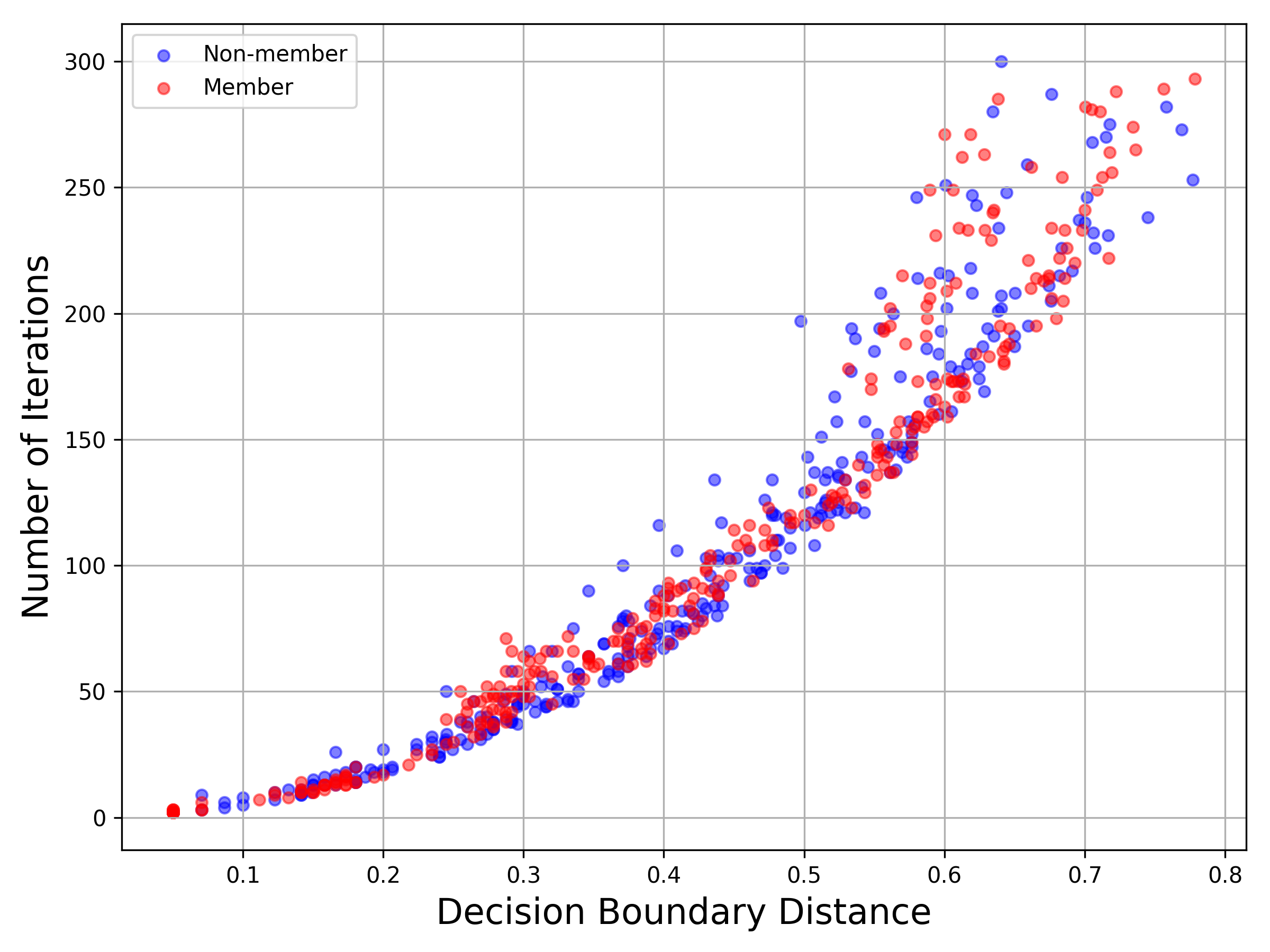}
    	}
    \end{minipage}
    \hfill
    \begin{minipage}[b]{0.48\linewidth}
        \centering
        \subfigure[b]{
    	    \includegraphics[width=\linewidth]{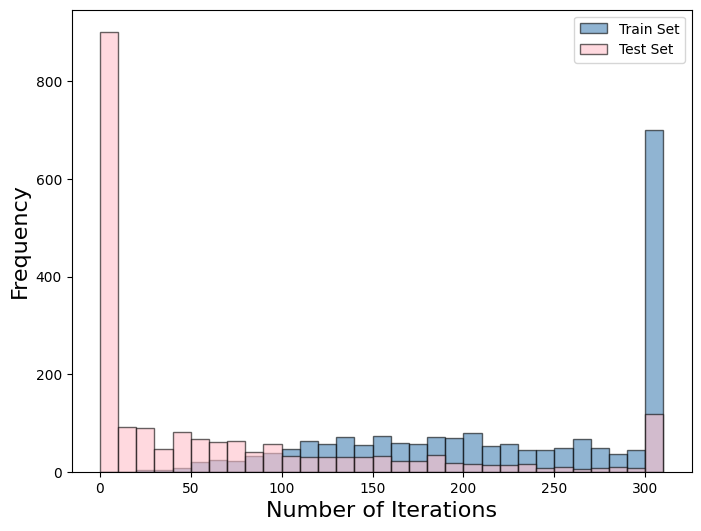}
    	}
    \end{minipage}
\caption{(a) Scatter diagram showing the relationship between the distance from samples to the decision boundary and the number of iterations required to generate adversarial examples using SimBA for ResNet trained on CIFAR10.
(b) Histogram about the number of iterations per-sample over 2k samples from the training set (blue) and the same number from the testing set. The adversarial samples are generated using SimBA for ResNet trained on CIFAR100.}
\label{fig:combined_mia_figures}
\vskip -0.2in
\end{figure*}

\section{Membership Inference Attack during Adversarial Examples Iteration}
In this section, we discuss the methodology of this paper, which aims to reveal privacy risks of the target model from the lens of adversarial samples' generation process.

\subsection{Motivation}
While prior MIA methods often rely on confidence scores or boundary distance \citep{choquette2021label,song2019privacy}, we find that boundary distance may not reliably distinguish members from non-members because different samples can have similar distance regardless of membership, as shown in Figure~\ref{fig:combined_mia_figures}(a). We plotted a scatter diagram showing the relationship between the distance from samples to the decision boundary and the number of iterations required to generate adversarial samples.  

Instead, we observe that the member samples generally require more iterations to generate adversarial samples than non-members. We use ``SimBA'' \citep{guo2019simple} to generate adversarial samples and record the number of iterations to generate its adversarial sample for each sample. The histogram is shown in Figure~\ref{fig:combined_mia_figures}(b): the samples from the training set (blue, member) need more iterations than those from the test set (pink, non-member). 

\begin{figure*}[ht]
\vskip 0.2in
\begin{center}
\centerline{\includegraphics[width=\columnwidth]{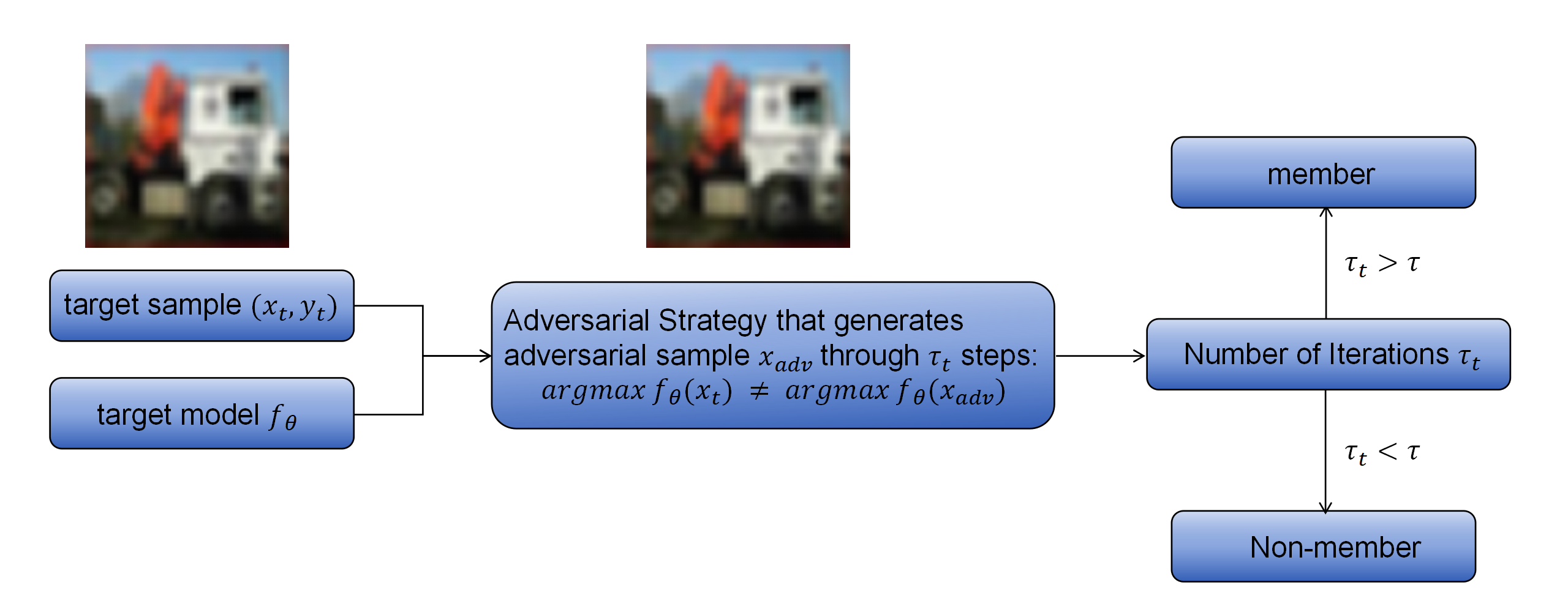}}
\caption{Diagrammatic sketch for \texttt{IMIA} to conduct MIA.}
\label{iteration-method}
\end{center}
\vskip -0.2in
\end{figure*}

This insight reveals a new, consistent signal for membership inference and motivates our method, IMIA, which leverages the number of iterations required to generate an adversarial sample for a target sample. 
If this number exceeds a preset threshold, this target sample will be inferred as a member; otherwise, as a non-member. The overall framework of \texttt{IMIA} is shown in Figure~\ref{iteration-method}. 

\subsection{Methodology} 
 Given the target model and the target images, the adversary can choose an adversarial strategy $\mathcal{S}$ in SimBA \citep{guo2019simple}, HopSkipJumpAttack \citep{chen2020hopskipjumpattack} and PGD \citep{madry2017towards} based on different MIA settings to generate adversarial samples and measure the number of iterations during this process. 

\paragraph{Score-based black-box attacks.} 
Adversary can obtain the full probability output of the target sample. As a result, we choose SimBA \citep{guo2019simple} to conduct adversarial attack which provides a simple but efficient strategy to change the target sample's output. The optimization goal in SimBA is to minimize the probability on the true label $y$ of the target sample $x$ :
\begin{align*}
    \min_{\delta} &\quad p_{f_{\theta}}(y|x+\delta)\\
    % \min_{\delta} &\quad \ell_{y}(x+\delta)\\
    \text{subject to:}&\quad\|\delta\|_2<d,\;\mathrm{queries}\leq M,
\end{align*}
where $\delta$ represents the perturbations and $M$ is the budget for the number of queries to the target model. SimBA solves this problem through randomly selecting a predefined orthonormal basis and either adding or subtracting it from the target sample according to the confidence scores which are checked if the target sample moves toward the decision boundary. During the process of generating adversarial samples for the target sample, we are concerned about the number of iterations after getting adversarial samples successfully.

\paragraph{Decision-based black-box attacks.} Different from score-based black-box attacks, the adversary can only obtain the label of the target input without any other information. In this case, we choose ``HopSkipJumpAttack'' \citep{chen2020hopskipjumpattack} whose performance is close to the white-box attack. Given the target image $(x,y)$, adversary starts from randomly choosing a point $x^{\prime}$ that is not classified to label $y$ by the target model and walks along the decision boundary to minimize the distance between the original image $x$ to its adversarial image $x^{\prime}$. In our methodology, we measure the number of iterations that adversarial samples can satisfy our request. 

\paragraph{White-box attacks.}  
In the white-box setting, We choose Projected Gradient Descent (PGD) \citep{madry2017towards} to generate adversarial samples. Formally, adversarial sample $x_{adv}$ is generated by:
\begin{align*}
    x_{adv}=\text{Clip}_{x, \epsilon} ( x_{N}^{adv} + \alpha \, \text{sign} \left( \nabla_{x} J(x_{N+1}^{adv}, y) \right)
\end{align*}
In \texttt{IMIA}, we will choose one of them to generate adversarial samples according to different MIA settings. The pseudocode for \texttt{IMIA} is listed in  \cref{alg:imia}. As shown in \cref{alg:imia}, given a target model $f_\theta$, a target sample $(x_t, y_t)$, an adversarial attack strategy $S$, and a threshold $\tau$, IMIA first computes the number of iterations $\tau_t$ required by $S$ to generate a successful adversarial example. The adversarial strategy $S$ is chosen based on the attack setting including PGD, SimBA and HopSkipJumpAttack as we described before. If $\tau_t \geq \tau$, the sample is classified as a \textit{member}; otherwise, it is classified as a \textit{non-member}.

In this paper, we consider three different attack strategies for different MIA settings, and we demonstrate the implementation details for each adversarial strategy in Appendix~\ref{appendix:experiment-details}. Note that we do not propose a novel method for adversarial attacks, instead, we only care about how the adversarial sample is generated from the original target sample.

\begin{algorithm}[tb]
   \caption{IMIA}
   \label{alg:imia}
   \begin{algorithmic}[1]
   \Require Target model $f_{\theta}$, target sample $(x_{\mathrm{t}}, y_{\mathrm{t}})$, adversarial strategy $\mathcal{S}$, threshold $\tau$
   \State ${\tau}_{\mathrm{t}} \gets \mathcal{S}(f_{\theta}, (x_{\mathrm{t}}, y_{\mathrm{t}}))$
   \Ensure Number of iterations ${\tau}_{\mathrm{t}}$ for generating adversarial sample
   \If{${\tau}_{\mathrm{t}} \geq \tau$}
       \State \Return $\mathds{1}({\tau}_{\mathrm{t}} \geq \tau)$ \Comment{Classify as member}
   \Else
       \State \Return $0$ \Comment{Classify as non-member}
   \EndIf
   \end{algorithmic}
\end{algorithm}

\section{Evaluation}
In this section, we will evaluate the effectiveness and universality of our algorithm. 
\subsection{Experiment Setup}
\textbf{Datasets.} In our experiment, we consider three different datasets which are all common in image recognition tasks. \textbf{CIFAR10} and \textbf{CIFAR100} all include 60k images and can be split into 10 and 100 classes respectively. 
In  PyTorch, CIFAR10 and CIFAR100 are divided into 50k images in the training dataset and others are in the test dataset. 
\textbf{STL-10} includes 5k images in the training set and 8k images in its test set. In our experiment, we use all samples from the training dataset to train the target model for each model architecture, and samples in the test set which do not participate in the training process are used to validate the model's accuracy.

\textbf{Target Model.} In our experiments, we use four different model architectures: ResNet50, VGG19, ResNeXt29\_2x64d, and DenseNet121. These models represent a diverse set of machine learning architectures. We train each model for 100 epochs in every dataset during the training process to ensure that the models are sufficiently trained and can produce meaningful outputs for membership inference attack. And we use the test set of the target model to verify the accuracy of the target model avoiding the low accuracy of the target model on the test set.

\textbf{Metrics.} We use a balanced evaluation set to evaluate our method and report the inference accuracy, AUROC scores, and the false positive rate (fpr) under different true positive rate (tpr). The inference accuracy considers both the true positive rate and the false positive rate and gives 50\% if the adversary guesses randomly. Area Under the Receiver Operating Characteristic Curve (AUROC) is the area under ROC curve, which is obtained by plotting the ratio of TPR to FPR at different thresholds. The closer the AUROC value is to 1, the better the attack performance.

In a balanced evaluation set, there is an equal number of member samples and non-member samples of the target model. As a result, our evaluation set consists of 3k samples from the target model's training set as member samples and uniformly selects the same number of samples from the test set as non-member samples. We repeat this many times to get different combinations of evaluation sets and finally report results on average.

In our paper, IMIA is a metric-based universal attack method which does not need any training data or shadow models, so to ensure a fair comparison, we also choose the metric-based method as our baselines instead of these methods using shadow models. The specific reasons can be found in the Appendix~\ref{appendix:reason}. Moreover, we display the specific time cost of computation to run IMIA and compare this to the cost of these methods based on shadow models in Appendix~\ref{appendix:cost}.

\begin{table*}[ht]
\caption 
{Membership inference results for different score-based attacks on CIFAR10, CIFAR100, and STL10 datasets in the black-box settings. We choose ``SimBA'' to generate adversarial samples for measuring the number of iterations for target samples. The evaluation set is composed of 3k samples from the training set and 3k samples from the test set of the target model, and we repeat this procedure 20 times. The inference accuracy(\%) and AUROC(\%) are reported in the average value with standard deviation. The best results in each case are in bold.}
\label{table 1}
% \belowrulesep=0pt
% \aboverulesep=0pt
% \vskip 0.15in
\begin{center}
\centering
\begin{small}
% \begin{sc}
\resizebox{\linewidth}{!}{
\begin{tabular}{ccccccccc}
\toprule
\multirow{2}{*}{Strategy}&\multicolumn{2}{c}{ResNet}&\multicolumn{2}{c}{ResNeXt}&\multicolumn{2}{c}{VGG}&\multicolumn{2}{c}{DenseNet} \\
\cmidrule(r){2-9}
&AUROC $\uparrow$ & Accuracy $\uparrow$ & AUROC & Accuracy & AUROC & Accuracy & AUROC & Accuracy \\
\midrule
\multicolumn{1}{l}{\textbf{CIFAR100}} & & & & & & & &\\
Softmax             & 88.91$\pm$ 0.13& 84.57$\pm$ 0.15& 63.66$\pm$ 0.09& 60.42$\pm$ 0.10& 63.73$\pm$ 0.12& 59.94$\pm$ 0.18& 67.70$\pm$ 0.11& 63.26$\pm$ 0.14 \\
Entropy             & 88.97$\pm$ 0.08& 84.59$\pm$ 0.11& 64.22$\pm$ 0.13& 54.42$\pm$ 0.07& 64.95$\pm$ 0.14& 61.26$\pm$ 0.12& 68.78$\pm$ 0.16& 64.00$\pm$ 0.11 \\
Mentr.              & \textbf{89.24$\pm$ 0.15}& \textbf{85.43$\pm$ 0.16}& 68.42$\pm$ 0.10& 64.48$\pm$ 0.12& \textbf{69.61$\pm$ 0.14}& \textbf{65.41$\pm$ 0.13}& 73.19$\pm$ 0.12& 67.90$\pm$ 0.18\\
IMIA(ours)  & 89.11$\pm$ 0.29& 83.62$\pm$ 0.29& \textbf{68.93$\pm$ 0.58}& \textbf{64.82$\pm$ 0.55}& 67.36$\pm$ 0.64& 64.68$\pm$ 0.55& \textbf{73.80$\pm$ 0.48}& \textbf{68.91$\pm$ 0.40}\\
\midrule
\multicolumn{1}{l}{\textbf{CIFAR10}} & & & & & & & &\\
Softmax             & 70.97$\pm$ 0.05& 65.84$\pm$ 0.15& 74.02$\pm$ 0.13& 68.05$\pm$ 0.19& 64.15$\pm$ 0.16& 60.60$\pm$ 0.15& 69.21$\pm$ 0.15& 64.17$\pm$ 0.14 \\
Entropy             & 71.05$\pm$ 0.13& 65.87$\pm$ 0.29& \textbf{74.35$\pm$ 0.14}& \textbf{68.24$\pm$ 0.17}& 64.39$\pm$ 0.15& 60.70$\pm$ 0.12& 69.51$\pm$ 0.16& 64.27$\pm$ 0.14 \\
Mentr.              & 71.95$\pm$ 0.12& 66.17$\pm$ 0.16& 73.54$\pm$ 0.15& 67.95$\pm$ 0.11& 64.56$\pm$ 0.12& 60.58$\pm$ 0.14& 69.44$\pm$ 0.15& 64.23$\pm$ 0.12\\
IMIA(ours)   & \textbf{74.43$\pm$ 0.15}& \textbf{68.35$\pm$ 0.13}& 73.34$\pm$ 0.09& 67.77$\pm$ 0.15& \textbf{66.53$\pm$ 0.13}& \textbf{62.87$\pm$ 0.15}& \textbf{75.20$\pm$ 0.09}& \textbf{68.97$\pm$ 0.15}\\
\midrule
\multicolumn{1}{l}{\textbf{STL10}} & & & & & & & &\\
Softmax             & 56.49$\pm$ 0.13& 55.53$\pm$ 0.15& 61.68$\pm$ 0.17& 59.43$\pm$ 0.11& 57.06$\pm$ 0.16& 56.63$\pm$ 0.12& 72.97$\pm$ 0.12& 68.52$\pm$ 0.13 \\
Entropy             & 55.91$\pm$ 0.18& 55.28$\pm$ 0.19& 62.08$\pm$ 0.12& 59.72$\pm$ 0.11& 57.75$\pm$ 0.14& 57.61$\pm$ 0.13& 73.22$\pm$ 0.13& 68.59$\pm$ 0.15 \\
Mentr.              & 61.05$\pm$ 0.11& 59.02$\pm$ 0.13& \textbf{70.84$\pm$ 0.18}& \textbf{66.37$\pm$ 0.10}& \textbf{62.68$\pm$ 0.14}& \textbf{59.92$\pm$ 0.12}& 78.41$\pm$ 0.11& 73.97$\pm$ 0.09\\
IMIA(ours)  & \textbf{61.26$\pm$ 0.15}& \textbf{59.70$\pm$ 0.12}& 69.25$\pm$ 0.15& 66.35$\pm$ 0.14& 61.32$\pm$ 0.08& 59.87$\pm$ 0.14& \textbf{81.18$\pm$ 0.12}& \textbf{75.41$\pm$ 0.11}\\
\bottomrule
\end{tabular}
}
% \end{sc}
\end{small}
\end{center}
% \vskip -0.1in
\end{table*}

\subsection{Results}
We analyze our results in the black-box and the white-box settings separately. In the black-box setting, there are two types: one is score-based attack where the target model outputs the confidence and labels at the same time; the other is the target model that only outputs hard labels. In the white-box setting, adversaries have access to the loss during the process of generating adversarial samples. Specifically, \texttt{IMIA} resorts to generate adversarial samples like ``PGD'' \citep{madry2017towards,tramer2017space} for white-box MIA attack, ``SimBA'' \citep{guo2019simple} for black-box MIA attack and ``HopSkipJumpAttack'' for hard labels MIA attack. 

\subsubsection{score-based setting}
In the score-based setting, the target model outputs the confidence and labels at the same time. In this case, we choose ``SimBA'' \citep{guo2019simple} as our strategy to generate corresponding adversarial samples of the original samples. Softmax Response \citep{song2019privacy}, Prediction Entropy \citep{salem2018ml} and Modified Entropy \citep{song2021systematic} as typical representatives in the score-based metric membership inference attack, we choose these as our baselines.

\begin{figure*}[ht]
\vskip 0.2in
    \begin{minipage}[b]{0.24\linewidth}
        \centering
        \subfigure[ResNet-CIFAR100]{
    	    \includegraphics[width=\linewidth]{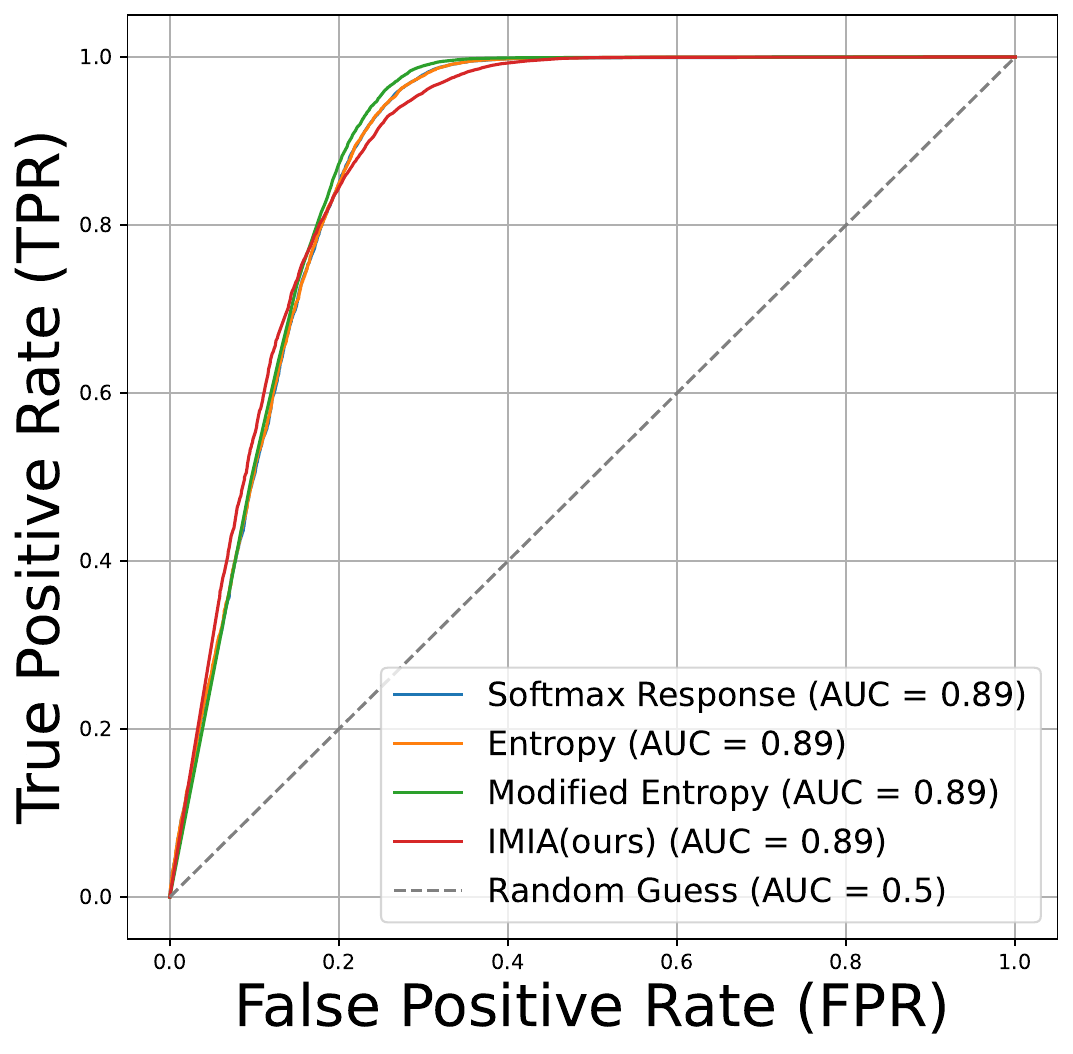}
    	}
    \end{minipage}
    \begin{minipage}[b]{0.24\linewidth}
        \centering
        \subfigure[ResNeXt-CIFAR100]{
    	    \includegraphics[width=\linewidth]{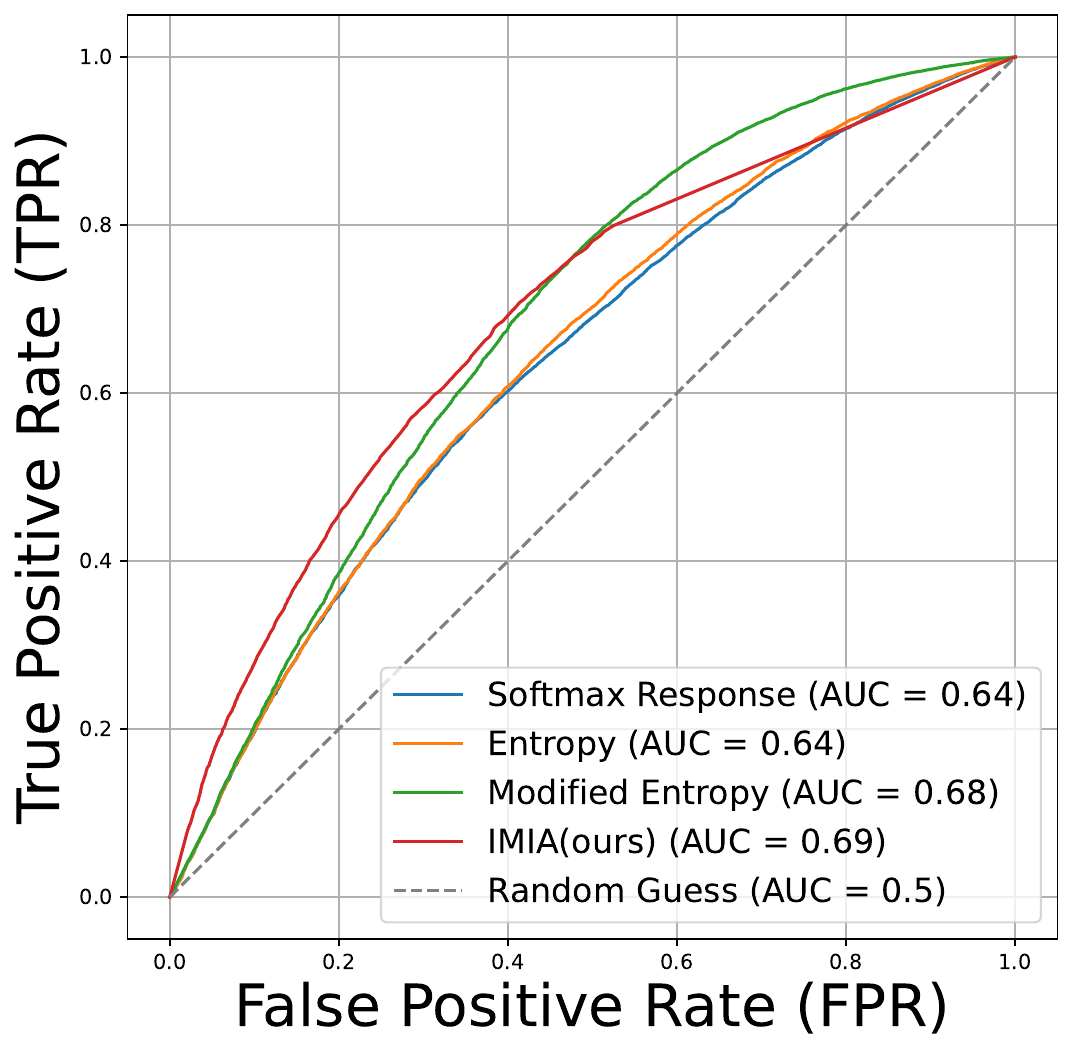}
    	}
    \end{minipage}
    \begin{minipage}[b]{0.24\linewidth}
        \centering
        \subfigure[VGG-CIFAR100]{
    	    \includegraphics[width=\linewidth]{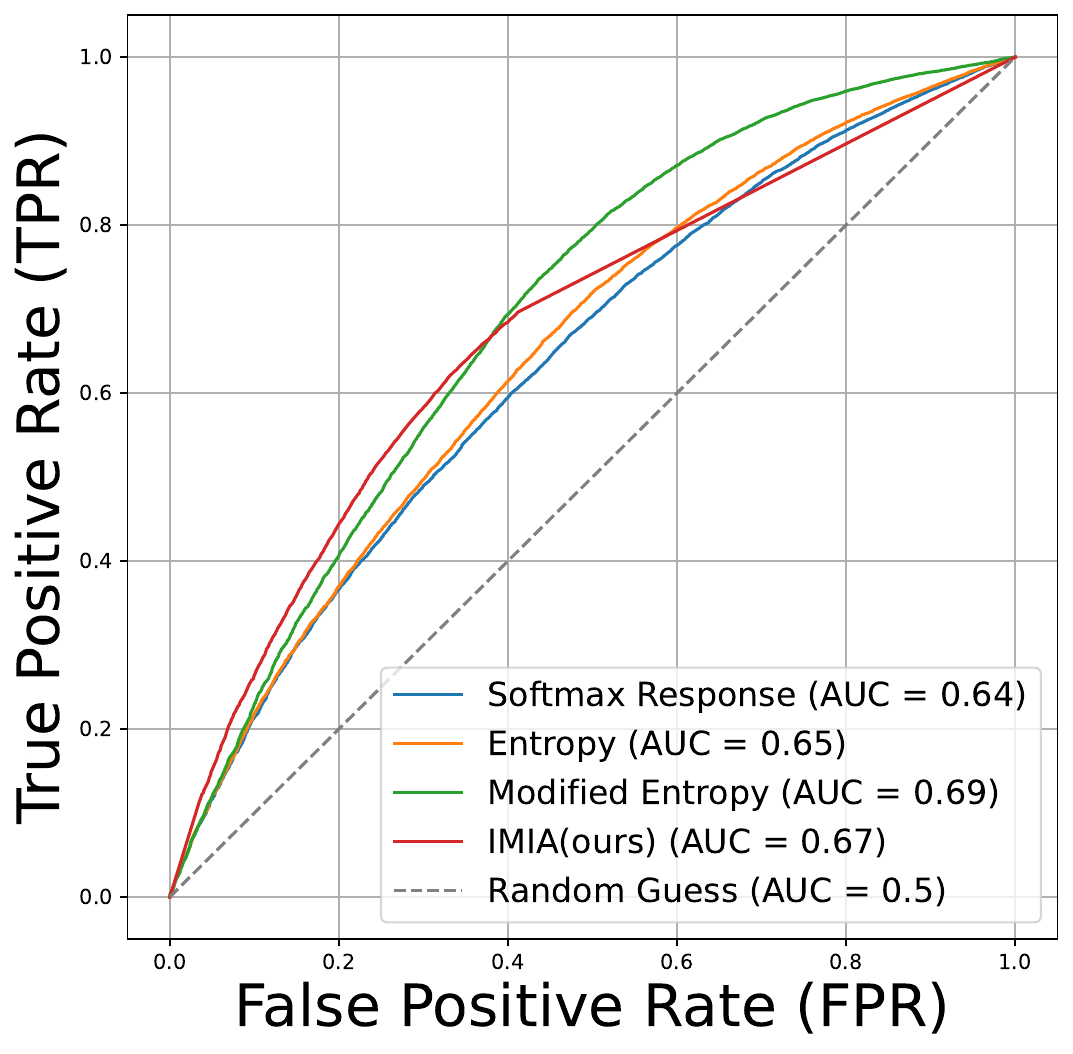}
    	}
    \end{minipage}
    \begin{minipage}[b]{0.24\linewidth}
        \centering
        \subfigure[DenseNet-CIFAR100]{
    	    \includegraphics[width=\linewidth]{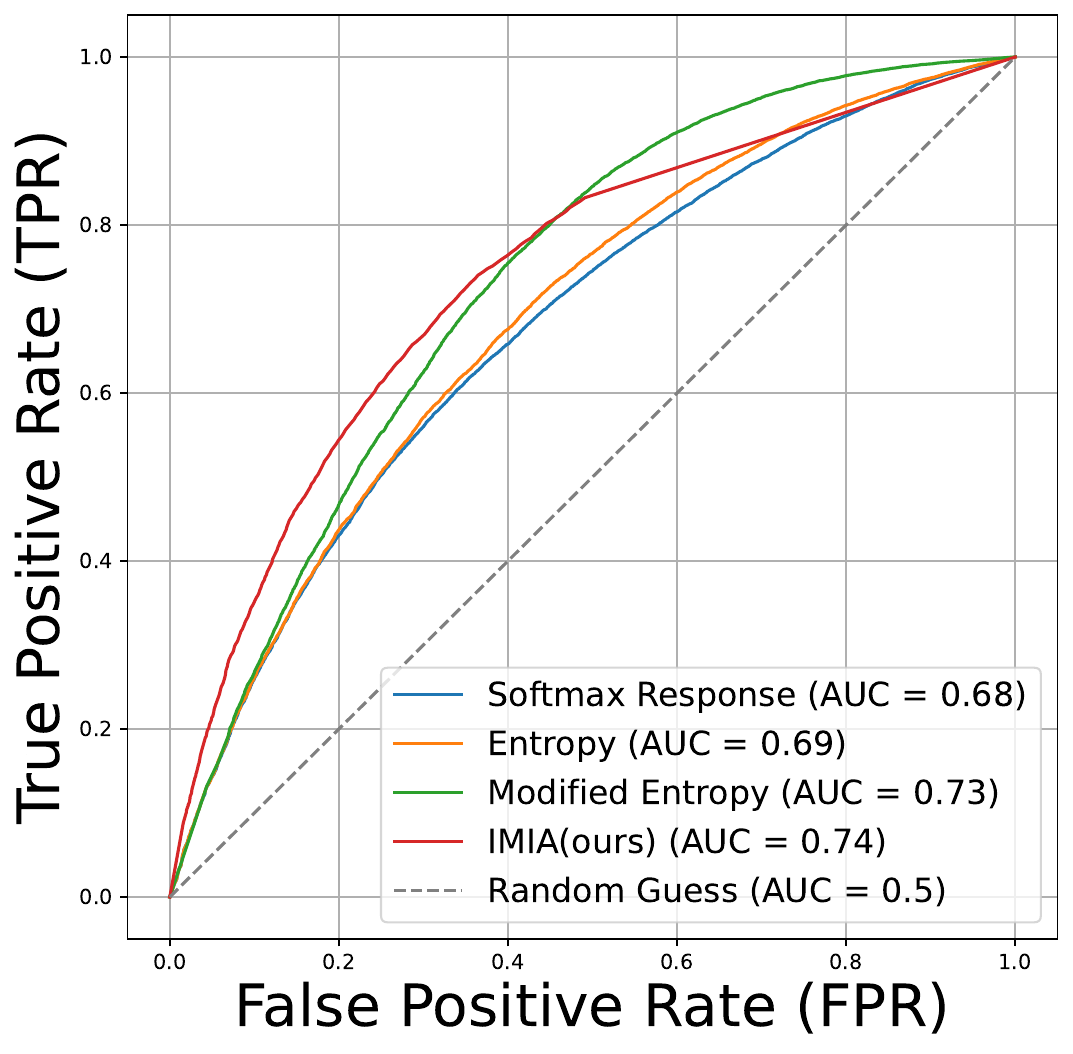}
    	}
    \end{minipage}
    \begin{minipage}[b]{0.24\linewidth}
        \centering
        \subfigure[ResNet-CIFAR10]{
    	    \includegraphics[width=\linewidth]{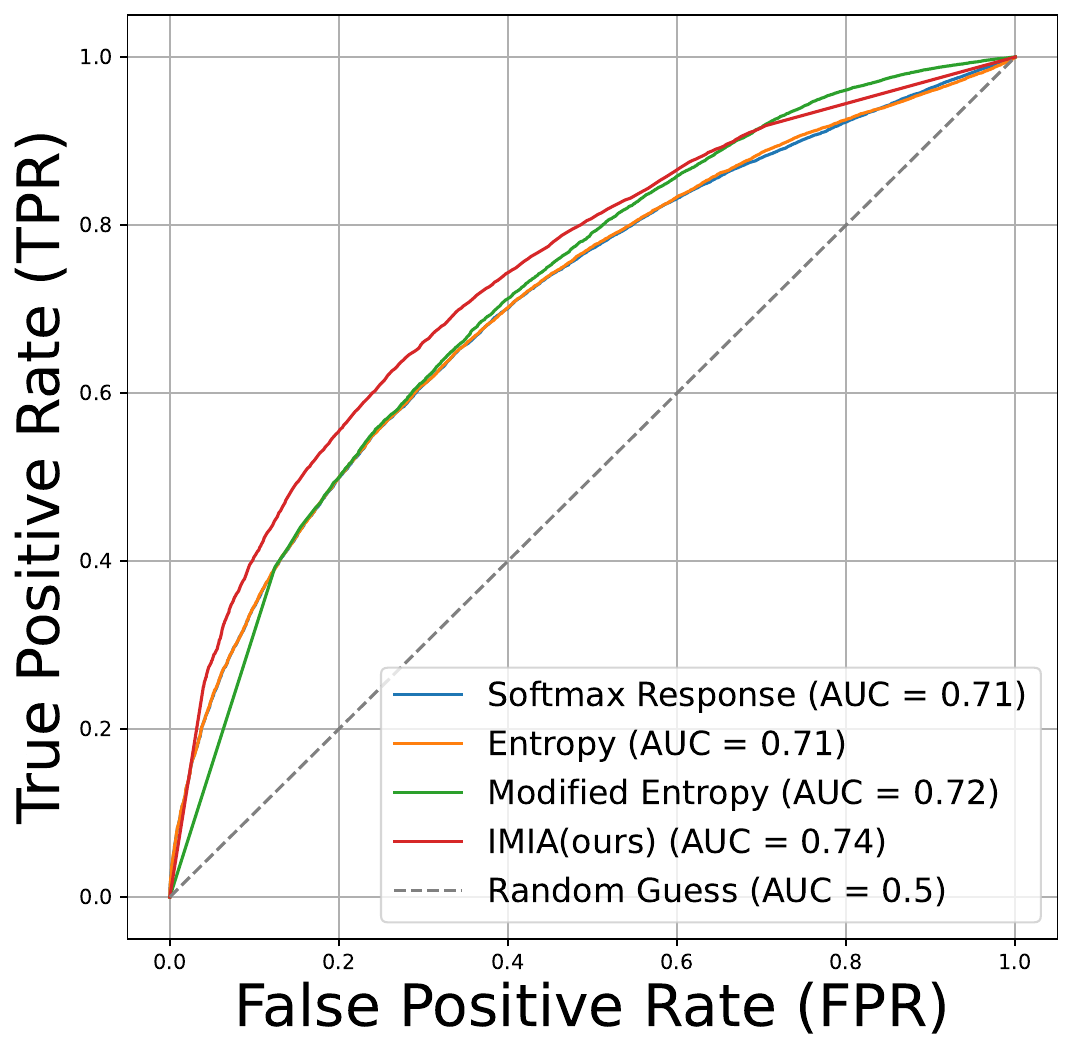}
    	}
    \end{minipage}
    \begin{minipage}[b]{0.24\linewidth}
        \centering
        \subfigure[ResNeXt-CIFAR10]{
    	    \includegraphics[width=\linewidth]{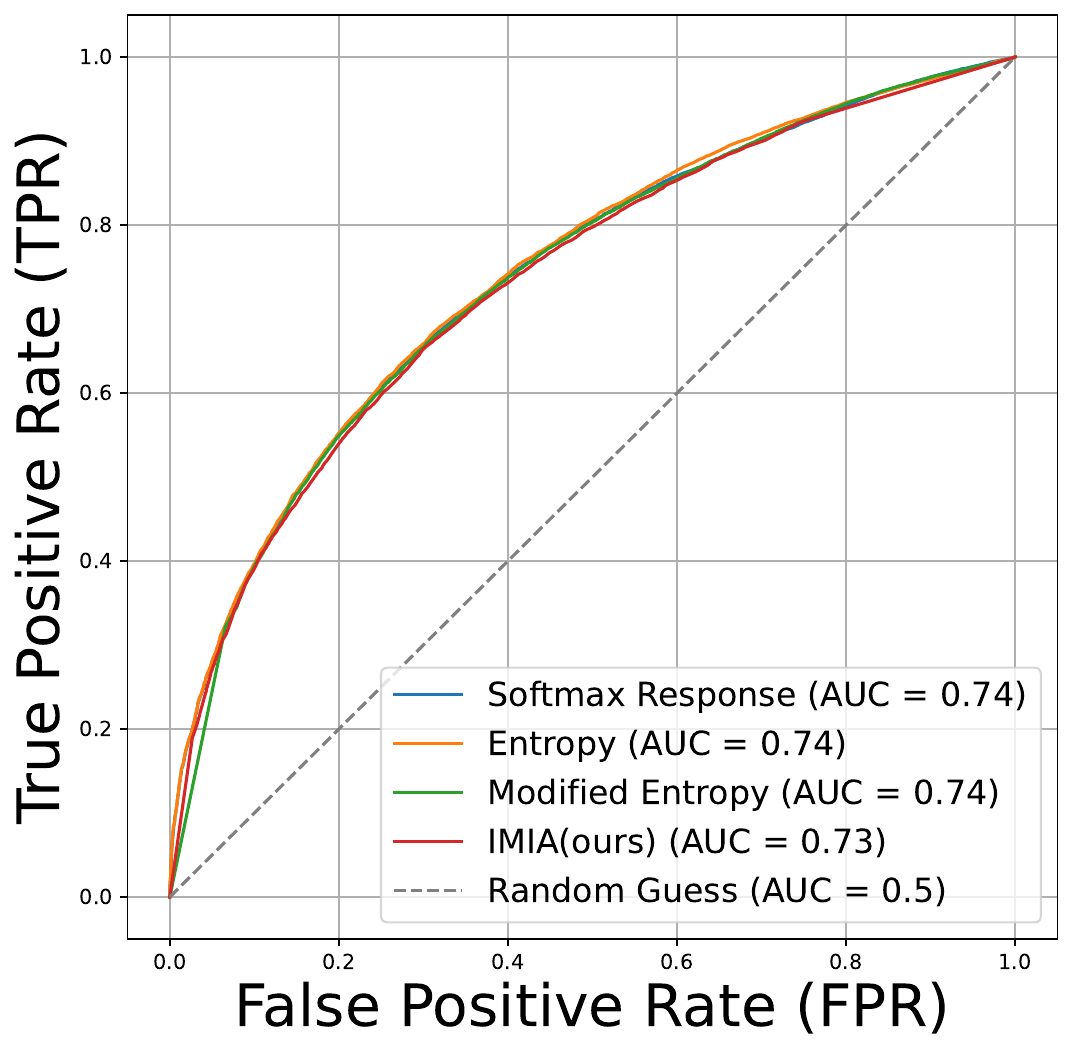}
    	}
    \end{minipage}
    \begin{minipage}[b]{0.24\linewidth}
        \centering
        \subfigure[VGG-CIFAR10]{
    	    \includegraphics[width=\linewidth]{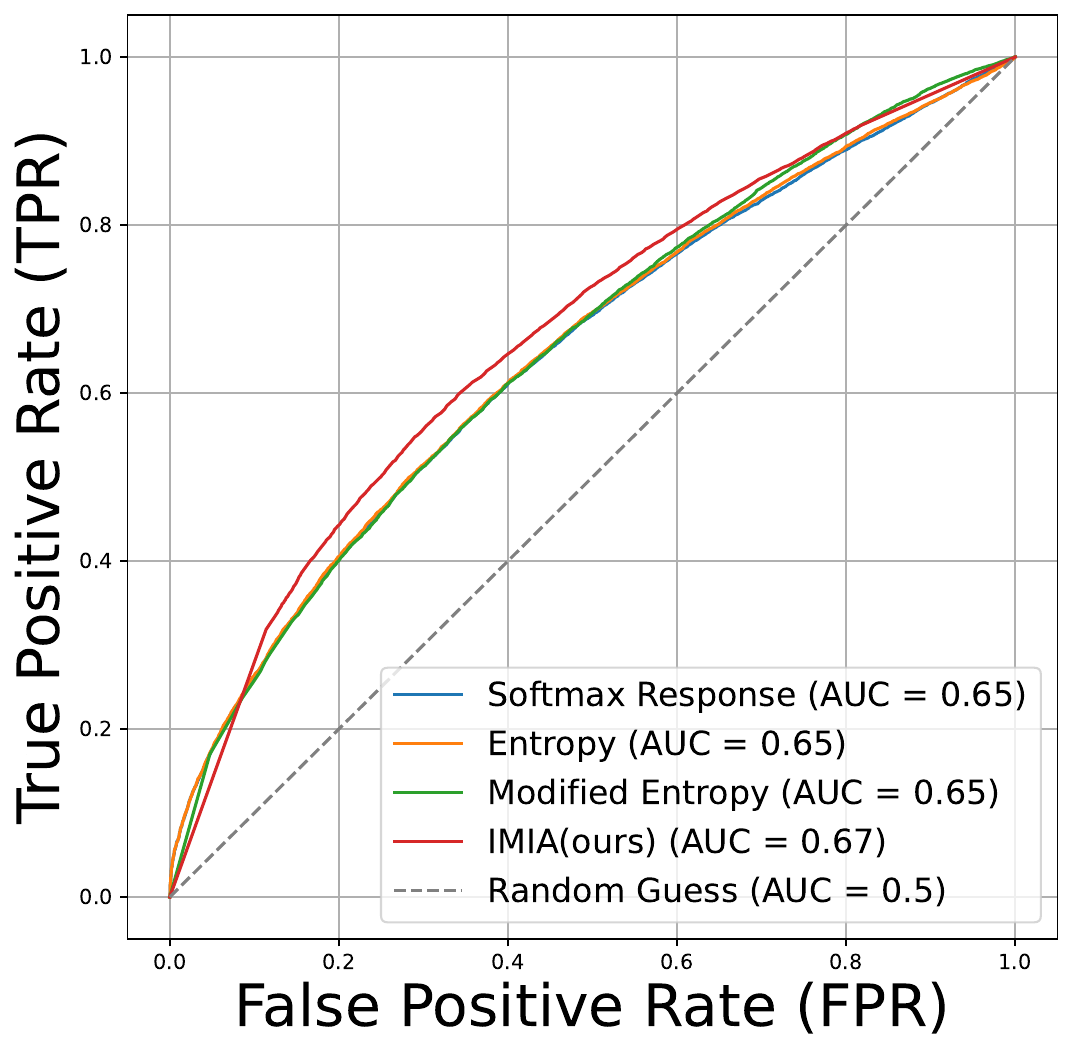}
    	}
    \end{minipage}
    \begin{minipage}[b]{0.24\linewidth}
        \centering
        \subfigure[DenseNet-CIFAR10]{
    	    \includegraphics[width=\linewidth]{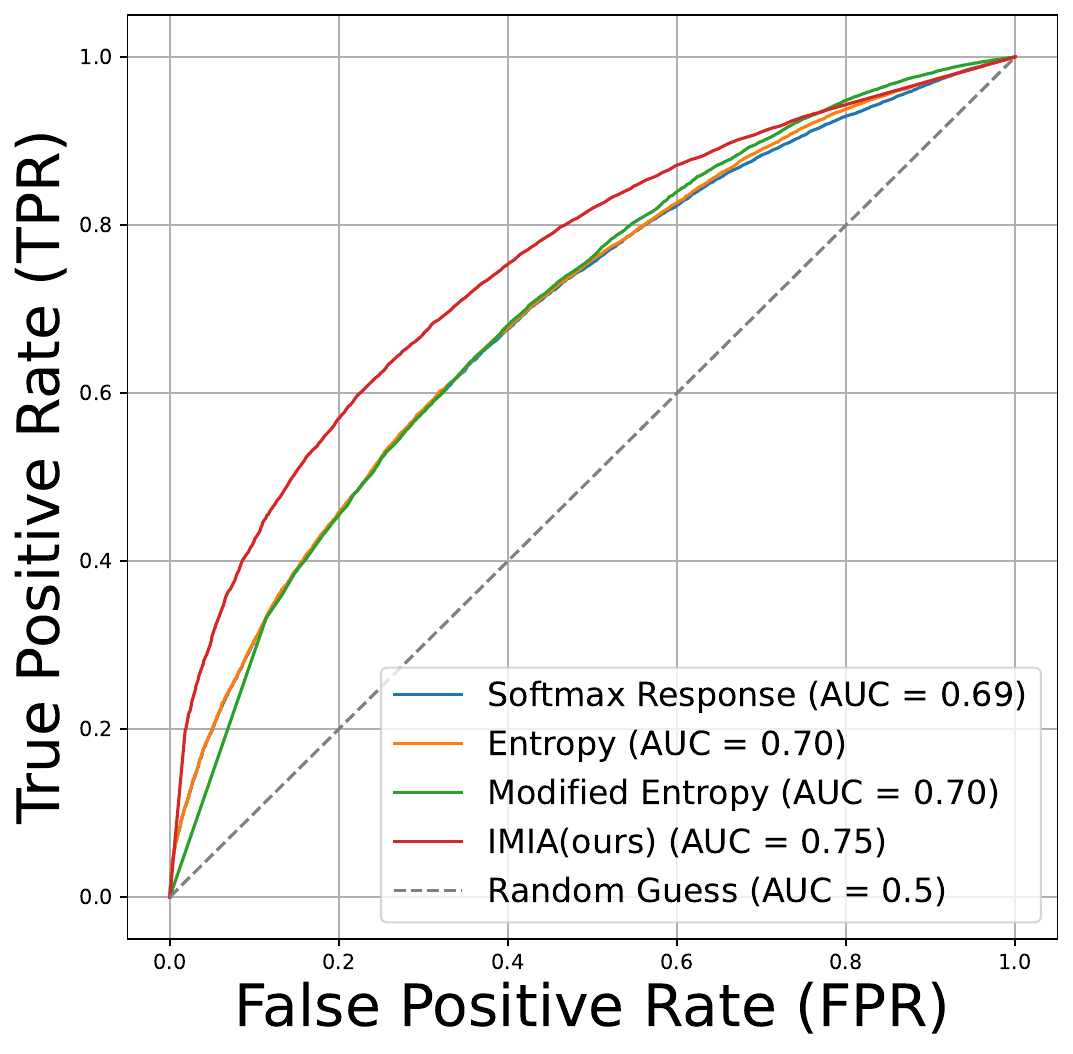}
    	}
    \end{minipage}
    \begin{minipage}[c]{0.24\linewidth}
        \centering
        \subfigure[ResNet-STL10]{
    	    \includegraphics[width=\linewidth]{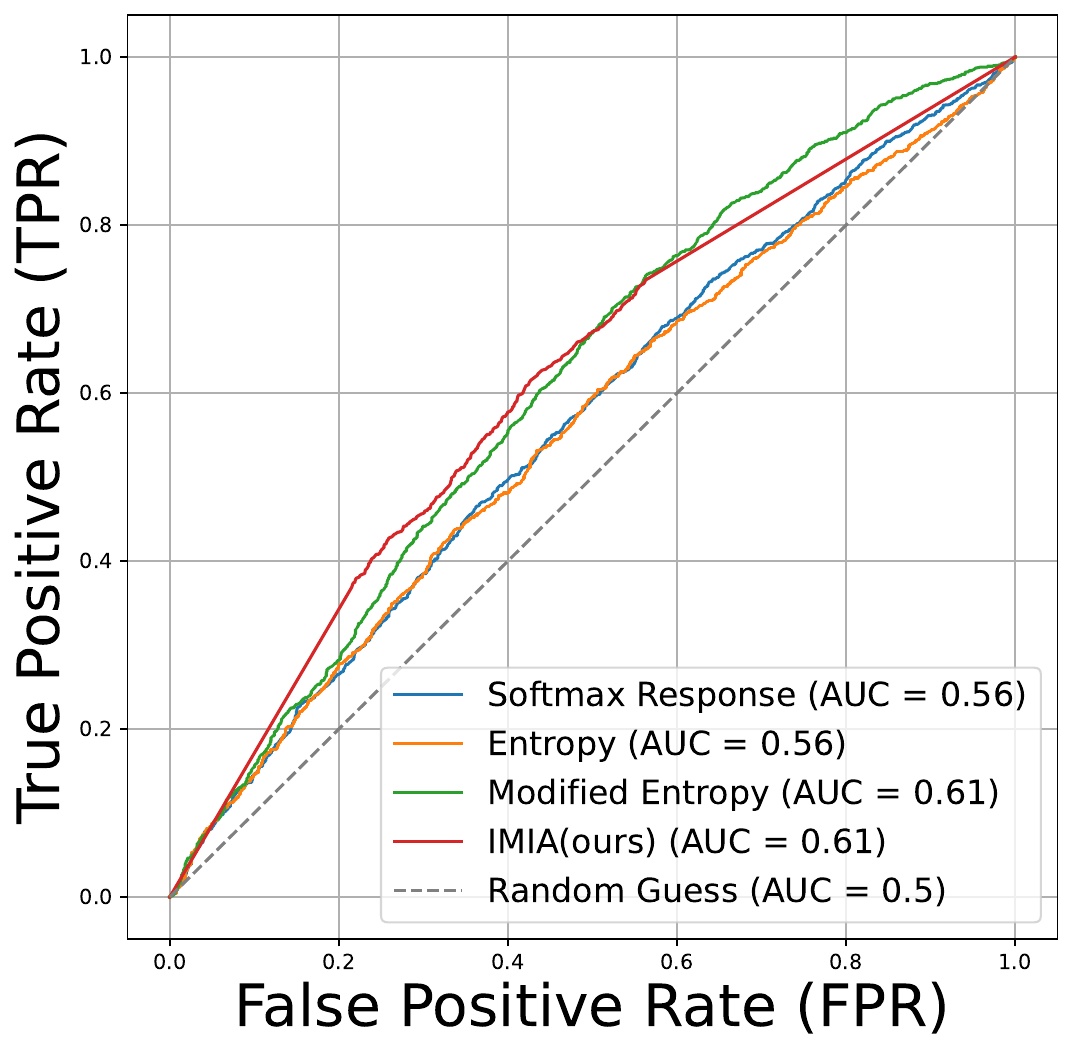}
    	}
    \end{minipage}
    \begin{minipage}[c]{0.24\linewidth}
        \centering
        \subfigure[ResNeXt-STL10]{
    	    \includegraphics[width=\linewidth]{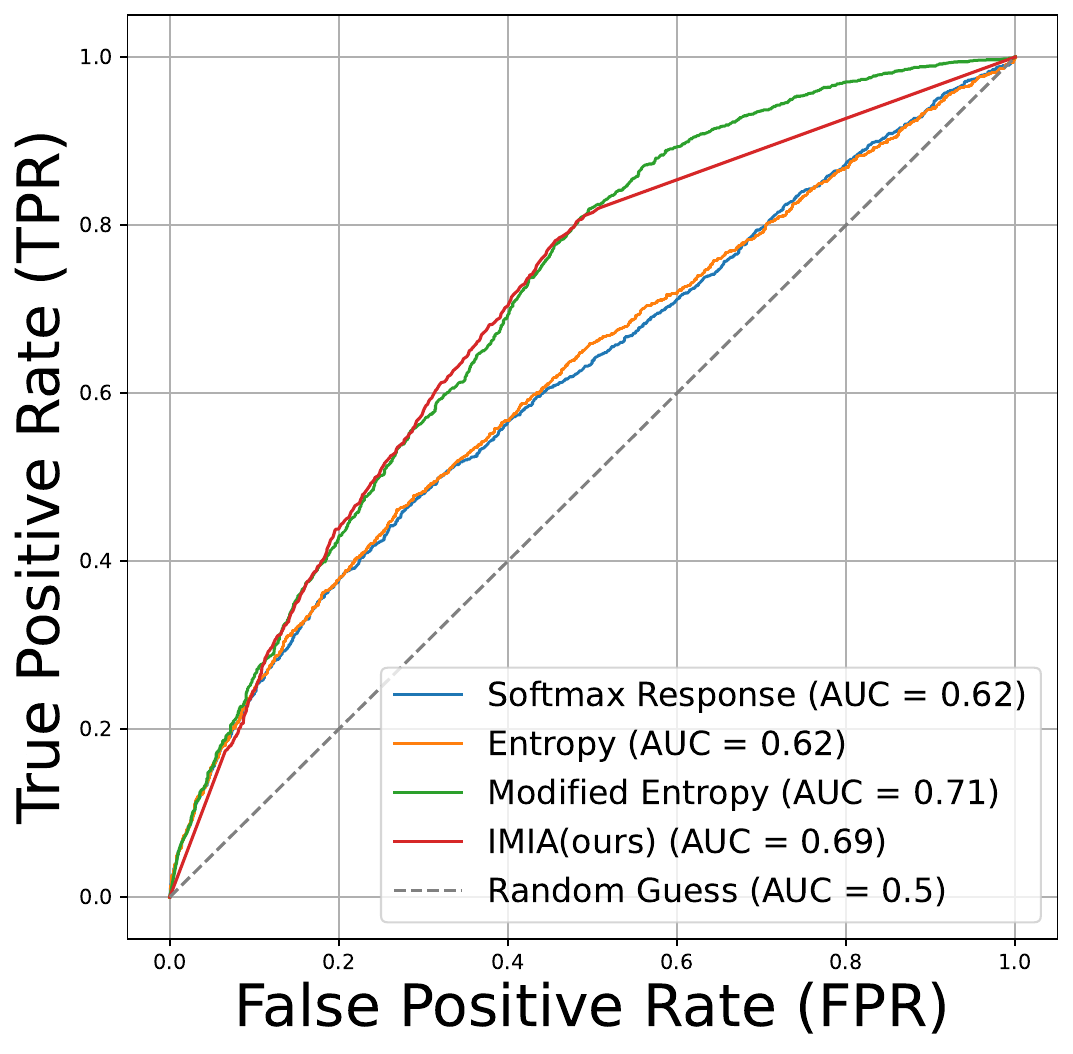}
    	}
    \end{minipage}
    \begin{minipage}[c]{0.24\linewidth}
        \centering
        \subfigure[VGG-STL10]{
    	    \includegraphics[width=\linewidth]{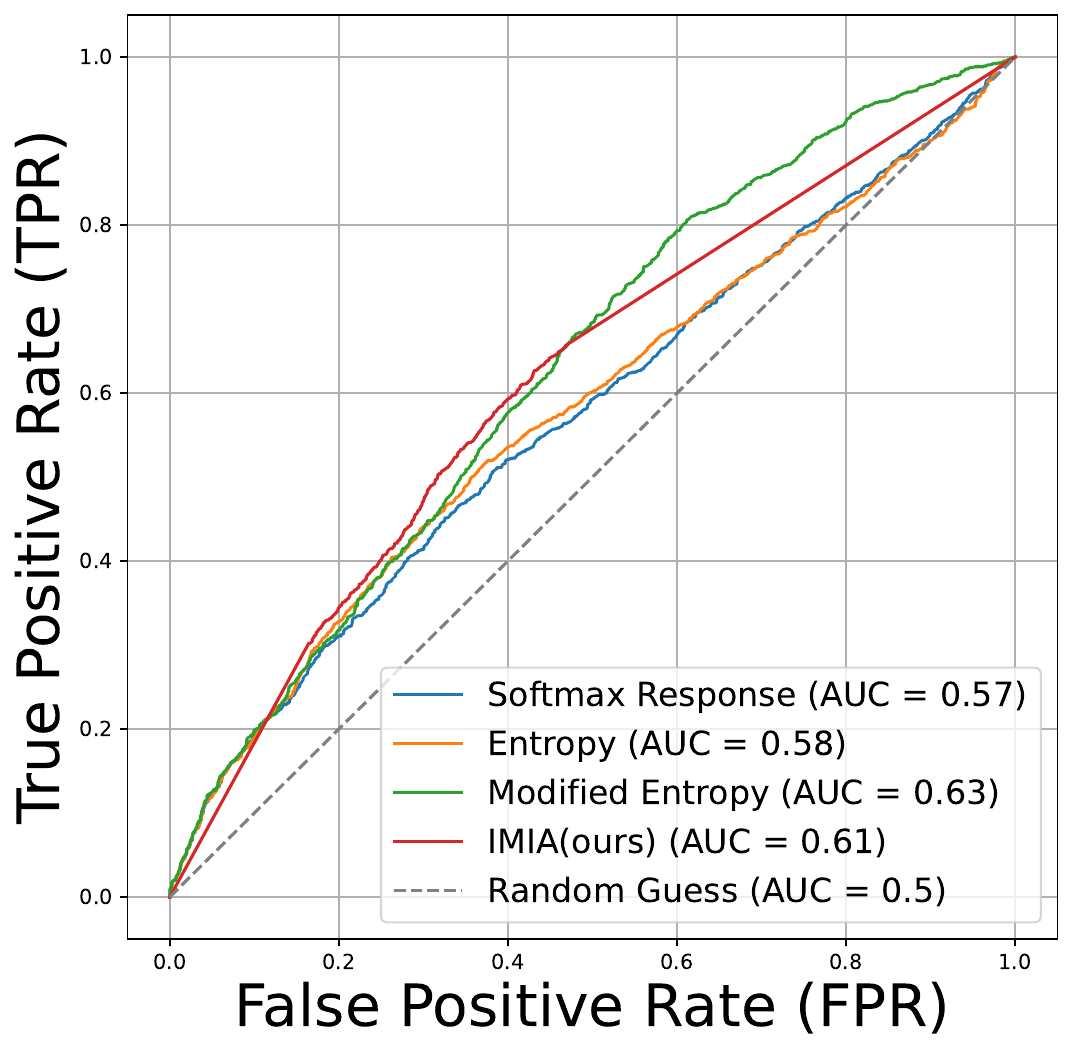}
    	}
    \end{minipage}
    \begin{minipage}[c]{0.24\linewidth}
        \centering
            \subfigure[DenseNet-STL10]{
    	    \includegraphics[width=\linewidth]{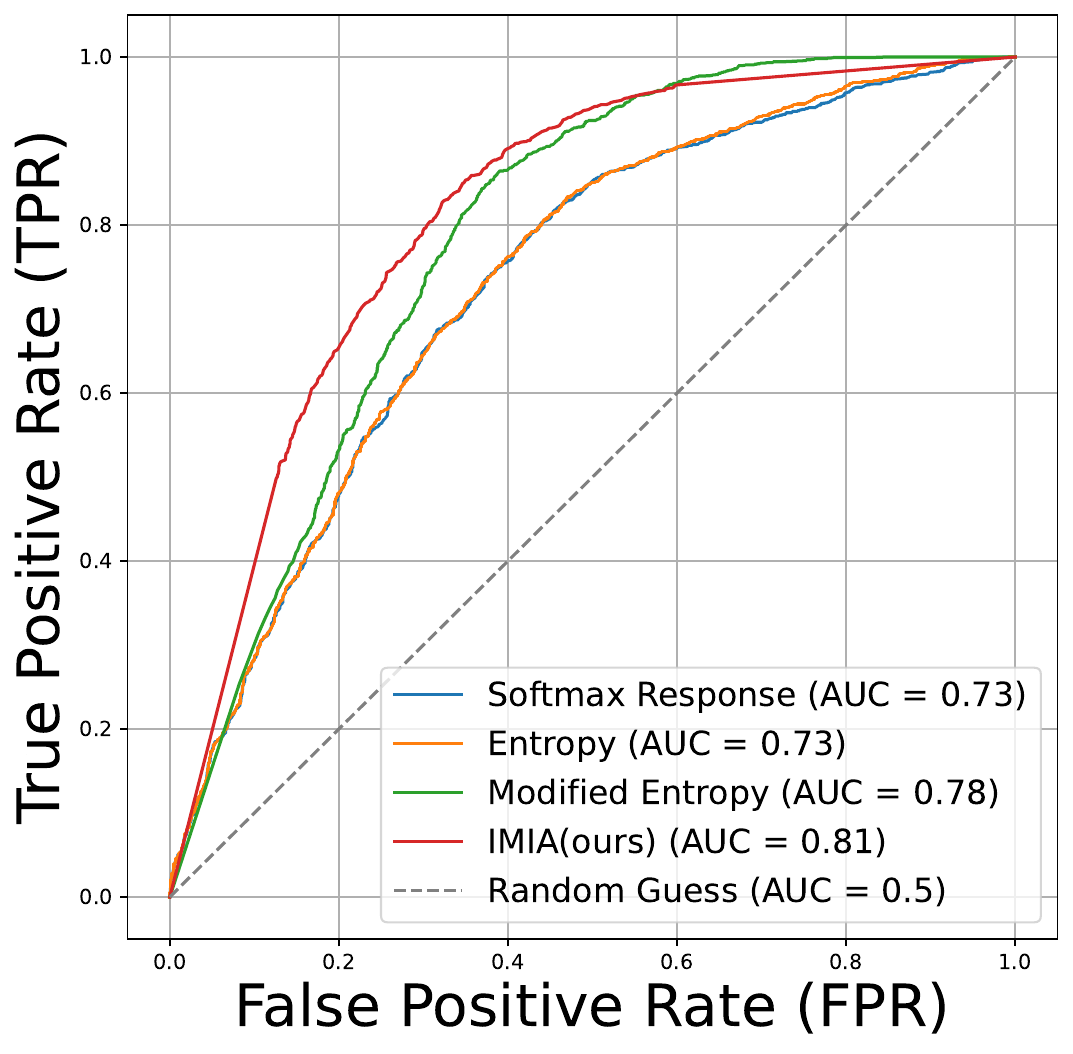}
    	}
    \end{minipage}
\caption{ROC curve on MIA for the combination of different models on CIFAR100, CIFAR10 and STL-10. They are drawn on the balanced evaluation set and correspond to \cref{table 1}.}
\label{tpr-fpr}
\vskip -0.2in
\end{figure*}

Table \ref{table 1} shows the results of our attacks and comparisons between \texttt{IMIA} and other baselines in the score-based setting. The results show our strategy that depending on the number of iterations performs well in distinguishing member samples and non-member samples. For example, for our proposed attack against CIFAR10 DenseNet classifier, the membership inference AUROC is increased from 69.21 to 75.20 on average and the accuracy is increased from 64.17 to 68.97. In other words, our strategy can effectively reveal the privacy risk of the target model during the process of generating adversarial samples. \cref{tpr-fpr} show the false positive rate under different true positive rate corresponding to Table \ref{table 1}. The figures illustrate how the false positive rate varies as the true positive rate changes. Our proposed attack has a relatively lower fpr than others.

\begin{table*}[t]
% \belowrulesep=0pt
% \aboverulesep=0pt
% \vskip 0.15in
\caption 
{Membership inference results for different decision-based attacks that only output hard labels on CIFAR10, CIFAR100 and STL10 datasets. We choose ``HopSkipJumpAttack'' strategy, and measure the distance and the number of iterations between original samples and adversarial samples. The inference accuracy(\%) and AUROC(\%) are reported in the average value with standard deviation. The best results in each case are in bold.}
\label{table 2}
\begin{center}
\centering
\begin{small}
% \begin{sc}
\resizebox{\linewidth}{!}{
\begin{tabular}{ccccccccc}
\toprule
\multirow{2}{*}{Strategy}&\multicolumn{2}{c}{ResNet}&\multicolumn{2}{c}{ResNeXt}&\multicolumn{2}{c}{VGG}&\multicolumn{2}{c}{DenseNet} \\
\cmidrule(r){2-9}
&AUROC $\uparrow$ & Accuracy $\uparrow$ & AUROC & Accuracy & AUROC & Accuracy & AUROC & Accuracy \\
\midrule
\multicolumn{1}{l}{\textbf{CIFAR100}} & & & & & & & &\\
Boundary            & \textbf{86.12$\pm$0.35 }& \textbf{81.59$\pm$0.43 }& \textbf{69.49$\pm$0.29} & \textbf{66.90$\pm$0.34 }& 63.29$\pm$0.33 & 61.90$\pm$0.26& 69.72$\pm$0.32 & 66.90$\pm$0.33 \\
IMIA(ours)    & 85.95$\pm$0.27 & 81.39$\pm$0.25 & 69.20$\pm$0.21 & 66.70$\pm$0.25 & \textbf{63.39$\pm$0.31} & \textbf{62.39$\pm$0.30}& \textbf{70.03$\pm$0.22 }& \textbf{67.20$\pm$0.22}\\
\midrule
\multicolumn{1}{l}{\textbf{CIFAR10}} & & & & & & & &\\
Boundary             & 72.84$\pm$0.30 & 66.70$\pm$0.25 & \textbf{70.43$\pm$0.26} &\textbf{ 65.60$\pm$0.31} & 66.50$\pm$0.23 & 63.0$\pm$0.21& 69.72$\pm$0.15 & 66.00$\pm$0.22 \\
IMIA(ours)    & \textbf{73.12$\pm$0.25} & \textbf{67.39$\pm$0.27} & 69.77$\pm$0.25 & 65.10$\pm$0.23 & \textbf{69.31$\pm$0.21} & \textbf{65.10$\pm$0.21}& \textbf{71.81$\pm$0.19}&\textbf{ 66.90$\pm$0.17}\\
\midrule
\multicolumn{1}{l}{\textbf{STL10}} & & & & & & & &\\
Boundary             & 61.45$\pm$0.27 & 60.24$\pm$0.31 & 70.82$\pm$0.35 & 67.34$\pm$0.31 & 63.78$\pm$0.24 & 61.60$\pm$0.20 & 82.42$\pm$0.13& 75.29$\pm$0.17 \\
IMIA(ours)     & \textbf{62.00$\pm$0.19}& \textbf{60.41$\pm$0.22} & \textbf{72.21$\pm$0.25} & \textbf{69.38$\pm$0.20} & \textbf{63.93$\pm$0.18} & \textbf{61.63$\pm$0.24} & \textbf{83.36$\pm$0.19}& \textbf{76.57$\pm$0.21}\\
\bottomrule
\end{tabular}
}
% \end{sc}
\end{small}
\end{center}
% \vskip -0.1in
\end{table*}

\begin{table*}[t]
\caption 
{Membership inference results for Loss and our method on CIFAR10, CIFAR100, and STL10 in the white-box setting. We choose ``PGD'' to generate adversarial samples and for the loss, we choose the cross entropy loss for each target sample. The inference accuracy(\%) and AUROC(\%) are reported in the average value with standard deviation. The best results in each case are in bold.}
\label{table 3}
% \belowrulesep=0pt
% \aboverulesep=0pt
% \vskip 0.15in
\begin{center}
\centering
\begin{small}
% \begin{sc}
\resizebox{\linewidth}{!}{
\begin{tabular}{ccccccccc}
\toprule
\multirow{2}{*}{Strategy}&\multicolumn{2}{c}{ResNet}&\multicolumn{2}{c}{ResNeXt}&\multicolumn{2}{c}{VGG}&\multicolumn{2}{c}{DenseNet} \\
\cmidrule(r){2-9}
&AUROC $\uparrow$ & Accuracy $\uparrow$ & AUROC & Accuracy & AUROC & Accuracy & AUROC & Accuracy \\
\midrule
\multicolumn{1}{l}{\textbf{CIFAR100}} & & & & & & & &\\
Loss              & 89.24$\pm$ 0.24& 85.68$\pm$ 0.26& 68.47$\pm$ 0.46& 65.14$\pm$ 0.51& 69.95$\pm$ 0.59& 65.87$\pm$ 0.60& 73.31$\pm$ 0.43& 68.42$\pm$ 0.46\\
Boundary          & 88.37$\pm$ 0.31& 82.82$\pm$ 0.26& 66.16$\pm$0.45 &63.97$\pm$0.49 &64.69$\pm$0.50 &62.62$\pm$0.51 &66.43 $\pm$0.40&64.46$\pm$0.43\\
IMIA(ours)  & \textbf{96.12$\pm$ 0.21}& \textbf{90.54$\pm$ 0.28}& \textbf{69.31$\pm$ 0.54}& \textbf{65.82$\pm$ 0.51}& \textbf{71.47$\pm$ 0.45}& \textbf{68.55$\pm$ 0.42}& \textbf{73.41$\pm$ 0.40}& \textbf{68.91$\pm$ 0.39}\\
\midrule
\multicolumn{1}{l}{\textbf{CIFAR10}} & & & & & & & &\\
Loss              & 72.26$\pm$ 0.49& 66.61$\pm$ 0.45& 74.33$\pm$ 0.35& 68.96$\pm$ 0.39& 65.09$\pm$ 0.36& 61.72$\pm$ 0.44& 69.94$\pm$ 0.35& 64.79$\pm$ 0.32\\
Boundary          & \textbf{75.45$\pm$0.47} & \textbf{69.03$\pm$0.45} &74.19$\pm$0.37 &68.62$\pm$0.37 &65.94$\pm$0.40 &61.67$\pm$0.42 &75.69$\pm$0.33 &69.82$\pm$0.29\\
IMIA(ours)  & 74.49$\pm$ 0.46& 68.97$\pm$ 0.39& \textbf{74.45$\pm$ 0.34}& \textbf{69.83$\pm$ 0.31}& \textbf{66.96$\pm$ 0.35}& \textbf{62.67$\pm$ 0.32}& \textbf{76.29$\pm$ 0.33}& \textbf{70.15$\pm$ 0.31}\\
\midrule
\multicolumn{1}{l}{\textbf{STL10}} & & & & & & & &\\
Loss              & 61.52$\pm$ 0.45& 59.20$\pm$ 0.33& \textbf{71.01$\pm$ 0.45}& \textbf{66.39$\pm$ 0.42}& \textbf{62.94$\pm$ 0.49}& \textbf{59.68$\pm$ 0.36}& 78.57$\pm$ 0.46& 74.06$\pm$ 0.45\\
Boundary          &60.29$\pm$0.43 &58.75$\pm$0.32 &68.38$\pm$0.50 &65.25$\pm$0.47 &59.66$\pm$0.45 &59.05$\pm$0.35 &81.78$\pm$0.46 &75.63$\pm$0.47\\
IMIA(ours)  & \textbf{61.94$\pm$ 0.47}& \textbf{59.64$\pm$ 0.36}& 70.07$\pm$ 0.53& 66.19$\pm$ 0.51& 61.46$\pm$ 0.52& 59.61$\pm$ 0.49& \textbf{82.19$\pm$ 0.61}& \textbf{75.63$\pm$ 0.50}\\
\bottomrule
\end{tabular}
}
% \end{sc}
\end{small}
\end{center}
% \vskip -0.1in
\end{table*}

\subsubsection{decision-based setting}
In another black-box setting called decision-based attack, the target model only outputs hard labels. 
In this case, we use ``HopSkipJumpAttack'' \citep{chen2020hopskipjumpattack} strategy to generate adversarial samples. 
We then compare our attack with the `` Decision boundary'' method \citep{choquette2021label} which has strong performance when the target model only outputs hard labels. 
The ``Boundary'' measures the distance from the adversarial samples to their decision boundary. Our \texttt{IMIA} measures the number of iterations during the generation of adversarial samples.
The comparison results are shown in Table~\ref{table 2}.
We can observe that \texttt{IMIA} has advantages over  ``Boundary'' method. 
For example, the membership inference AUROC and accuracy in STL10 are all higher than ``Boundary'' method for all classifiers. 
Furthermore, for the VGG model on the CIFAR10, our \texttt{IMIA} achieves an accuracy about $2.8\%$ higher than ``Boundary''. 
Though \texttt{IMIA} is simple, in this most difficult situation, it can still work efficiently.

\subsubsection{white-box setting}
We also evaluate \texttt{IMIA} in the white-box setting and show the comparison in Table~\ref{table 3} where the best results in each case are in bold. In the white-box setting, ``Loss'' method \citep{yeom2018privacy} is one of the most common methods to measure the degree of privacy leaks, so we also use ``Loss'' method as baseline and compute the cross entropy loss to quantify the privacy risks associated with the target model. At the same time, we also compare the results of ``Decision Boundary'' method.
In this setting, we use ``PGD'' \citep{madry2017towards} methodology to generate adversarial samples. 
Table~\ref{table 3} shows that \texttt{IMIA}  surpasses the ``Loss'' method both in the inference accuracy and AUROC in most classifiers and datasets.
Especially, when applied to the DenseNet architecture on CIFAR10 dataset,  \texttt{IMIA} can achieve an accuracy over $5\%$ higher than the ``Loss'' method. 

\subsection{Summary}
In three different application conditions, the increasing inference accuracy and AUROC value prove that our methodology \texttt{IMIA} has great performance in distinguishing the member data and the non-member data. Even in the most difficult situation, our methodology can still work well. All results show that \texttt{IMIA} has great adaptive ability and can be applied in both white-box and black-box settings without knowing data from the training set. 
\section{Conclusion}
In this paper, We propose a universal membership inference attack method, called \texttt{IMIA}, which performs the membership inference attack from the perspective of adversarial samples' generation process. The key idea of \texttt{IMIA} is to measure the number of iterations by the process of generating the adversarial samples, and use this metric to infer whether the target samples belong to the model's training set or not.
\texttt{IMIA} with different adversarial strategies can be applied in different settings.
Accordingly, we conduct experiments in different MIA settings and on different datasets such as CIFAR10, CIFAR100 and STL10 datasets under different model architectures. 
Our experiment results show that our proposed methodology has great performance in different situations with higher AUROC values and inference accuracy compared to the other metric-based MIA algorithms. 
All experiments highlight the superior performance of \texttt{IMIA} and prove that \texttt{IMIA} is universal and adaptable in most settings to detect the privacy risk of the target model while requiring fewer computational resources, making it a more efficient choice for MIA.

\bibliographystyle{plainnat}   % 你也可以用 abbrvnat、unsrtnat 等
\bibliography{example_paper}   % 不要加 .bib 后缀，比如 myrefs.bib -> \bibliography{myrefs}

\newpage
\appendix
\onecolumn
\section{Implementation details for each adversarial strategy}
\label{appendix:experiment-details}
In Table~\ref{tab:simba_params}, we show the implementation details for the Simple Black-box Attack (SimBA) under the score-based setting. The parameters in our experiments are similar with \citet{guo2019simple}. The most difference is the perturbation size ($\epsilon$) which we set a small value for observing the number of iterations during the adversarial samples' generation process.

We implement the HopSkipJump Attack (HSJA) \citep{chen2020hopskipjumpattack} under a decision-based black-box setting. We follow the standard setup of the HopSkipJump Attack (HSJA) \citep{chen2020hopskipjumpattack}, using $L_2$ constraint with input clipping to $[0, 1]$, and set the number of iterations to 100. Other hyperparameters, such as step size search strategy and evaluation limits, are set according to default values or tuned empirically for stability in Table~\ref{tab:hsja_params}.

We employ the Projected Gradient Descent (PGD) attack \citep{madry2017towards} under the white-box setting with an $L_\infty$ perturbation budget of $\epsilon = \frac{3}{255}$, step size $\alpha = 0.001$, and 50 attack steps. A random start is enabled to initialize the attack from a perturbed point within the allowed norm ball. The attack is untargeted.

\begin{table}[h]
\centering
\caption{SimBA Attack Hyperparameters}
\label{tab:simba_params}
\begin{tabular}{ll}
\toprule
\textbf{Parameter} & \textbf{Value} \\
\midrule
Maximum iterations (\texttt{max\_iters}) & 300 \\
Frequency dimensions (\texttt{freq\_dims}) & 32 \\
Stride (\texttt{stride}) & 7 \\
Perturbation size ($\epsilon$) & 0.05 \\
$L_\infty$ bound (\texttt{linf\_bound}) & 0.0 (unbounded) \\
Perturbation order (\texttt{order}) & \texttt{rand} \\
Targeted attack (\texttt{targeted}) & False \\
Pixel attack (\texttt{pixel\_attack}) & False (frequency domain) \\
Logging interval (\texttt{log\_every}) & 40 \\
\bottomrule
\end{tabular}
\end{table}

\begin{table}[h]
\centering
\caption{HSJA Attack Hyperparameters}
\label{tab:hsja_params}
\begin{tabular}{ll}
\toprule
\textbf{Parameter} & \textbf{Value} \\
\midrule
Clipping Range (\texttt{clip\_min}, \texttt{clip\_max}) & [0, 1] \\
Norm Constraint (\texttt{constraint}) & $L_2$ \\
Number of Iterations (\texttt{num\_iterations}) & 100 \\
Binary Search Confidence (\texttt{gamma}) & 1.0 \\
Step Size Search Strategy (\texttt{stepsize\_search}) & Geometric progression \\
    Max Gradient Evaluations (\texttt{max\_num\_evals}) & 1e4 \\
Initial Gradient Evaluations (\texttt{init\_num\_evals}) & 100 \\
\bottomrule
\end{tabular}
\end{table}

\section{Justification for Not Comparing with Shadow Model-Based Attacks}
\label{appendix:reason}
As mentioned in our contributions, IMIA does not need any training data or train shadow models. But regarding the state-of-the-art MIA methods, whether it is LiRA\citep{carlini2022membership}, RMIA\citep{zarifzadeh2023low}, Oslo\citep{prashanth2024recite} or YOQO\citep{wuyou}, all of these methods require training additional models. In LiRA\citep{carlini2022membership}, RMIA\citep{zarifzadeh2023low}, and YOQO\citep{wuyou}, shadow models are trained, and in Oslo\citep{prashanth2024recite}, source and validation models are trained. Although some methods\citep{prashanth2024recite} can conduct attacks using a small number of shadow models, the fact is that they still train shadow models. Moreover, all methods based on training shadow models require large amounts of data, which is acknowledged in \citet{prashanth2024recite}. In our method, no additional models need to be trained, and there is no requirement to know training data to perform IMIA. Therefore, to ensure a fair comparison, we also choose these metric-based methods like Softmax and Entropy that do not require training shadow models as our baselines.

\section{The computational time cost of IMIA}
\label{appendix:cost}
In our paper, our IMIA method determines membership by counting the number of iterations required to generate adversarial samples. This process is carried out on a per-sample basis and, since generating adversarial samples in a black-box setting does not require storing gradient information—only forward propagation—the time spent on a single sample is relatively small. We list the specific time cost of computation for each adversarial strategy in Table~\ref{table 4}. For example, in the white-box setting, achieving an adversarial sample only needs 33.4 milliseconds under the VGG model trained with CIFAR100.
By contrast, methods that rely on training shadow models, even though the models can be reused, typically take at least half an hour to train a single shadow model on an RTX 4090 GPU. When the sample size is large, IMIA may take longer. However, our attack does not require a substantial amount of training data to be effective.

\begin{table*}[t]
\caption 
{The comparison results of computational time cost between IMIA and those methods using shadow models. Time reported under the attack settings in our paper refers to the generation of a single adversarial example per sample. Time reported for RMIA/LiRA refers to the training time of a single shadow model. We compute different adversarial strategies under different model structures and datasets.}
\label{table 4}
\centering
% \vskip 0.15in
% \begin{small}
\resizebox{\linewidth}{!}{
\begin{tabular}{llll}
\toprule
\textbf{Attack Setting} & \textbf{Attack Method} & \textbf{Model-Data} & \textbf{Time (ms/sample)} \\
\midrule
Score-based attack                & SimBA        & ResNet-CIFAR10       & 709.66 \\
Decision-based (label-only) attack & HSJA         & DenseNet-STL10       & 3767.1 \\
White-box attack                  & PGD          & VGG-CIFAR100         & 33.41 \\
\midrule
RMIA / LiRA                       & Shadow model & ResNet-CIFAR10       & 30min \\
\bottomrule
\end{tabular}
}

\end{table*}

%%%%%%%%%%%%%%%%%%%%%%%%%%%%%%%%%%%%%%%%%%%%%%%%%%%%%%%%%%%%

\appendix

 \end{document}